\documentclass[pdflatex,sn-mathphys-ay]{sn-jnl}

\usepackage{graphicx}%
\usepackage{multirow}%
\usepackage{amsmath,amssymb,amsfonts}%
\usepackage{amsthm}%
\usepackage{mathrsfs}%
\usepackage[title]{appendix}%
\usepackage{xcolor}%
\usepackage{textcomp}%
\usepackage{manyfoot}%
\usepackage{booktabs}%
\usepackage{algorithm}%
\usepackage{algorithmicx}%
\usepackage{algpseudocode}%
\usepackage{listings}%

\theoremstyle{thmstyleone}%
\theoremstyle{thmstyletwo}%

\theoremstyle{thmstylethree}%

\newcommand{\defn}{\ensuremath{\stackrel{\textrm{def}}{=}}}

\begin{document}

\title[Unravelling how winds and surface heat fluxes control the Atlantic Ocean's meridional heat transport]{Unravelling how winds and surface heat fluxes control the Atlantic Ocean's meridional heat transport}


\author*[1,2,3,4]{\fnm{Dhruv} \sur{Bhagtani}}\email{dhruv.bhagtani@anu.edu.au; dhruv.bhagtani@princeton.edu}

\author[1,2,5]{\fnm{Andrew} \sur{McC. Hogg}}\email{andy.hogg@anu.edu.au}

\author[6]{\fnm{Ryan} \sur{M. Holmes}}\email{ryan.holmes@bom.gov.au}
\author[5,7]{\fnm{Navid} \sur{C. Constantinou}}\email{navid.constantinou@unimelb.edu.au}


\affil[1]{\orgdiv{Research School of Earth Sciences}, \orgname{Australian National University}, \orgaddress{\city{Acton}, \state{ACT}, \country{Australia}}}

\affil[2]{\orgdiv{Australian Research Council Centre of Excellence for Climate Extremes}, \orgaddress{\country{Australia}}}

\affil[3]{\orgdiv{Department of Geosciences, Princeton University, Princeton, NJ}, \orgaddress{\country{USA}}}

\affil[4]{\orgdiv{High Meadows Environmental Institute, Princeton University, Princeton, NJ}, \orgaddress{\country{USA}}}

\affil[5]{\orgdiv{Australian Research Council Centre of Excellence for the Weather of the 21st Century}, \orgaddress{\country{Australia}}}

\affil[6]{\orgdiv{Australian Bureau of Meteorology}, \orgaddress{\city{Sydney}, \state{NSW}, \country{Australia}}}

\affil[7]{\orgdiv{School of Geography, Earth and Atmospheric Sciences}, \orgname{University of Melbourne}, \orgaddress{\city{Parkville}, \state{VIC}, \country{Australia}}}



\abstract{The North Atlantic Ocean circulation, fueled by winds and surface buoyancy fluxes, carries 1.25~PettaWatts of heat poleward in the subtropics, and helps in regulating global weather and climate patterns. 
Here, we assess the relative impacts of changes in winds and surface heat fluxes on the Atlantic Ocean circulation and heat transport on short timescales ($<$10 years) and long timescales ($>$50 years) using ocean simulations. 
We decompose the circulation and heat transport into warm and cold cells (resembling a subtropical gyre and the dense overturning circulation respectively), and a mixed cell capturing waters transitioning between warm and cold regions. 
Warm and mixed cells transport more heat poleward as wind stress increases; however, these anomalies are compensated by reductions in the cold cell's heat transport. 
Warm and cold cells transport more heat poleward when we increase meridional heat flux gradients. 
Our findings underscore the distinct roles of winds and surface heat fluxes in controlling the Atlantic Ocean’s meridional heat transport.}




\maketitle

\section{Introduction}\label{sec1}

\label{sec:New-intro}

The oceanic and atmospheric circulation together advect around 5.5$\,\mathrm{PW}$ (1$\,\mathrm{PW} \defn 10^{15}\,\mathrm{W}$) of heat poleward to regulate global weather and climate patterns \citep{Masuda1988}.
Modeling studies \citep{Msadek2013, Stepanov2016, Liu2022-sfc-heat-fluxes} and hydrographic surveys \citep{Johns2011, Trenberth2019, Williams2019} estimate that the large-scale North Atlantic Ocean circulation, driven by a combination of winds and surface buoyancy fluxes, transfers close to $1.25\,\mathrm{PW}$ heat poleward in the subtropics, which is about three-quarters of the total oceanic heat transport at these latitudes.
Future changes in the Atlantic Ocean's circulation and heat transport are expected due to climate change-induced variability in surface forcing \citep{Mecking2023}.
Such variations include the intensification of westerlies \citep{Shaw2024} and polar amplification \citep{Rantanen2022}, yet the relative importance of winds and surface heat fluxes in driving the Atlantic Ocean's meridional heat transport (MHT) is not fully understood.
The goal of this study is to better understand how winds and surface heat fluxes independently influence the Atlantic Ocean's MHT through their controls on the large-scale circulation.

The Atlantic Ocean circulation is dominated by two distinct (albeit interconnected) mechanisms. First, convection in the subpolar North Atlantic produces dense water that contributes to driving the deep arm (between 1000$\,\mathrm{m}$--3000$\,\mathrm{m}$) of the mid-depth overturning circulation, also referred to as the Atlantic Meridional Overturning Circulation (AMOC) \citep{Cessi2019}. 
The AMOC transfers heat northward in both hemispheres \citep{Bryan1991, Williams2019}. 
Additionally, a shallow (upper $\sim1000\,\mathrm{m}$) subtropical gyre in the North Atlantic confined between $20^{\circ}\mathrm{N}-40^{\circ}\mathrm{N}$ carries warm waters poleward via the Gulf Stream, with a relatively colder equatorward flow.
The subtropical gyre is believed to be primarily driven by wind stress \citep{Sverdrup1947}, though recent studies highlight an important control of surface buoyancy forcing \citep{Hogg2020, Bhagtani2023}.
Moreover, the subtropical gyre adjusts to changes in surface forcing on seasonal to interannual timescales, while the AMOC responds more slowly (e.g., see timescales of changes in the two circulations in the paper by \cite{Bhagtani2023}).
Because these two circulatory systems are modified by external forcing on different timescales, we separately examine their short-term ($<10\,\mathrm{years}$) and long-term ($>50\,\mathrm{years}$) adjustments. 
This approach elucidates key dynamical links between surface forcing -- such as winds and heat fluxes -- and the resulting patterns of heat transport.

The first attempts to isolate the contributions of the subtropical gyre and the AMOC to the heat transport used Eulerian averaging techniques \citep{Hall1982, Bryan1982}.
These techniques assume that the heat flux by the AMOC at a given latitude is the vertical integral of the product of a zonally averaged velocity and a zonally averaged temperature, with the residual being the gyre component of the MHT (assuming zero net volume transport at the given latitude). 
Based on these Eulerian-averaging techniques, \cite{Roemmich1985} and \cite{Bryden2001} designate the AMOC as the primary driver of Atlantic MHT by virtue of the large temperature difference between its warmer, shallow northward branch and the colder, deep equatorward branch.
However, flow across isopycnals is much weaker than along isopycnals \citep{Abernathey2022-oceanmixing}, therefore, a decomposition in depth space may not realistically separate the subtropical gyre and the AMOC.
Moreover, since the ocean's density is primarily controlled by temperature in our region of interest (that is, in the tropics and mid-latitudes; see \cite{Caneill2022}), a decomposition in either density or temperature space may better represent the physical pathways of these two circulatory features.

Several studies have used potential density or temperature to isolate the subtropical gyre and the AMOC.
\cite{Boccaletti2005} and \cite{Greatbatch2007} separated the circulation in temperature space and found similar magnitudes of MHT by the resulting subtropical gyre and the AMOC.
\cite{Talley2003} used density to divide the ocean into shallow, intermediate, and deep waters and found similar northward heat contributions of the shallow and deep water masses (corresponding to a gyre and an inter-hemispheric overturning, respectively), in line with \cite{Boccaletti2005} and \cite{Greatbatch2007}.
These partitioning techniques compute the advective component of the MHT.
In contrast, using output from numerical models alongside RAPID array observations, \cite{Xu2016} found that the subtropical gyre (isopycnal flow above a potential density referenced to $2000\,\mathrm{dbar}$ of $34.75\,\mathrm{kg}\,\mathrm{m}^{-3}$) carries a small net southward heat flux in density space.
Since the Gulf Stream is warmer than the corresponding return flow, it is unusual that the subtropical gyre defined by \cite{Xu2016} carries heat southward and suggests a limitation of defining the gyre in density space.
\cite{Jones2023} further speculate that this southward MHT may be because the Gulf Stream does not necessarily follow isopycnal contours.
Collectively, these studies underscore the sensitivity of the partitioning technique on the proportion of heat fluxes carried by the subtropical gyre and the AMOC and highlights the importance of adopting a framework that is physically consistent with the observed structure and pathways of ocean circulation.
Given the issues with partitioning in depth and density spaces as highlighted above, we instead decompose the circulation in temperature space (e.g., see \cite{Boccaletti2005}) to understand the role of warm and cold circulations (resembling the subtropical gyre and the AMOC respectively) in carrying heat poleward. 

The gyre and the AMOC have recently been shown to be coupled \citep{Berglund2022}, therefore, a complete delineation between the two features may not be correct \citep{Johns2023}.
This coupling is acknowledged in the decomposition by \cite{Ferrari2011}, who extend the latitude--temperature framework of \cite{Boccaletti2005} to include a mixed circulation that captures water masses transiting between warm (resembling the North Atlantic subtropical gyre) and cold (resembling the mid-depth overturning) temperatures. 
In this paper, we decompose the Atlantic Ocean circulation in latitude--temperature space into a warm, cold, and mixed cell following \cite{Ferrari2011} (section~\ref{sec:models_methods}).
We analyze a series of flux-forced simulations (described in section~\ref{sec:models_methods}) in which we completely separate the effects of wind stress and surface buoyancy forcing from each other.
These simulations differ from \cite{Ferrari2011} in that they are more suitable to answer our main research question: how does each surface forcing \textit{independently} contribute to the Atlantic Ocean's MHT?
We present our findings in section~\ref{sec:Results} and discuss their implications in section~\ref{sec:discussion}.

\section{Models and methods}
\label{sec:models_methods}

\subsection{Model setup}
\label{subsec:setup}

\begin{figure}[h!]
    \includegraphics[width=1.0\textwidth, trim = {0.25cm 2.7cm 0cm 3.3cm}, clip]{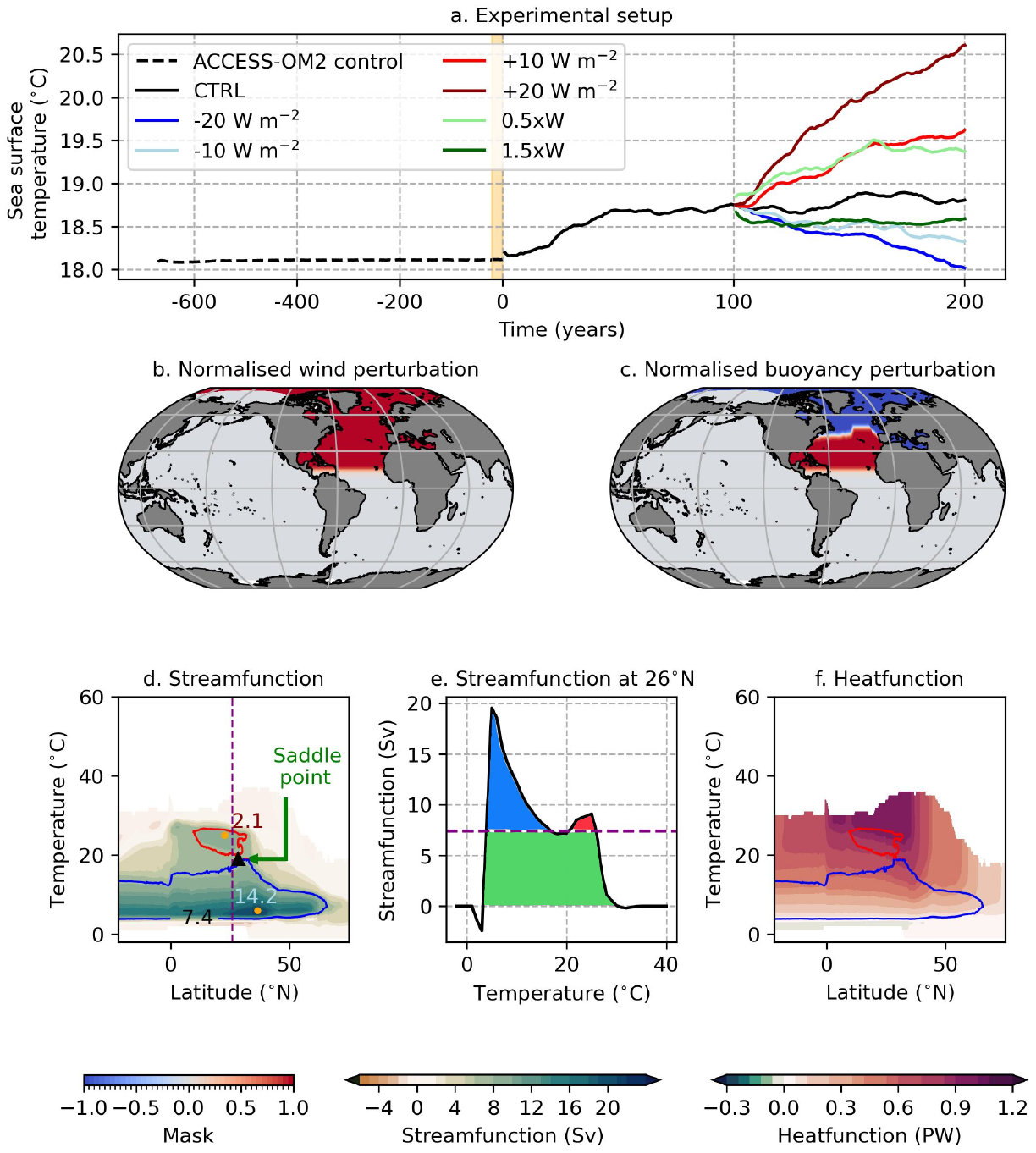}\vspace{-1em}
    \caption{(a).~Time series of the mean sea surface temperature for all experiments illustrating the model spinup.
             (b).~Normalised map for zonal and meridional surface wind stress perturbations. The perturbed winds stress is equal to the product of the CTRL's wind stress, the normalised anomaly map, and the desired multiplicative factor. We apply the CTRL's surface wind stress for regions outside the North Atlantic.
             (c).~Normalised map for surface heat flux perturbations.
             (d).~Streamfunction in latitude--temperature space calculated from years $101-110$ of the CTRL simulation highlighting the strength of the mixed (black text), warm (red text), and cold (blue text) cell. Vertical dashed line denotes 26$^{\circ}$N. The black solid triangle marks the saddle point temperature~$\theta_{\text{saddle}}$.
             (e).~Streamfunction for the 26$^{\circ}$N transect marking warm (red), cold (blue), and mixed (green) cells. The horizontal purple line denotes the upper bound of the mixed circulation strength ($=7.4\,\textrm{Sv}$).
             (f).~Heatfunction in latitude--temperature space obtained by cumulatively integrating the streamfunction in panel~(d).
            }
    \label{fig:Model_setup}
\end{figure}

We run a series of ocean simulations using the Modular Ocean Model~5 \citep{Griffies2012} (at 0.25$^{\circ}$ horizontal resolution with 50 unevenly spaced vertical levels) by applying climatological surface boundary fluxes of heat, freshwater, and momentum at a 3-hourly frequency.
We name these simulations as ``flux-forced simulations'' on account of prescribed surface forcing.
The boundary fluxes are computed from the last 20 years (orange bar in Fig.~\ref{fig:Model_setup}a) of a 670-year-long equilibrated ACCESS-OM2 ocean--sea ice model simulation \citep{Kiss2020}.
The ACCESS-OM2 model is also run at a 0.25$^{\circ}$ horizontal resolution using repeat year forcing based on the JRA55-do atmospheric reanalysis dataset \citep{Tsujino2018} and initialized from the World Ocean Atlas 2013 dataset \citep{Locarnini2013, Zweng2013}.
The effect of winds and surface buoyancy fluxes on the ocean is usually computed from bulk formulae (e.g., as done in ACCESS-OM2) and relies on the model's sea surface temperature and atmospheric variables.
By prescribing surface fluxes in our flux-forced simulations, we are able to modify each forcing independent of the other, allowing us to ascertain their relative contributions to the Atlantic Ocean's MHT.
The only remaining buoyancy contribution occurs from frazil ice formation that in reality occurs throughout the water column, hence we are physically motivated to model frazil dynamically in our flux-forced experiments instead of applying a surface flux to simulate its behavior.

The flux-forced control simulation is branched off from the 670-year-long ACCESS-OM2 simulation (black dashed line in Fig.~\ref{fig:Model_setup}a) and allowed to stabilize for 100 years, after which we branch off a series of flux-forced experiments with perturbations restricted to the North Atlantic basin.
We categorize these simulations into two types:~\emph{(i)}~multiplicative variations in wind stress, and \emph{(ii)}~additive variations in surface meridional heat flux gradients.
Since wind stress curl exerts a significant control on near-surface circulation, we multiply the North Atlantic Ocean's wind stress by a constant factor (Fig.~\ref{fig:Model_setup}b), which changes the curl by the same proportion.
We analyze two simulations where we change the wind stress in the North Atlantic by a factor of 0.5 and 1.5 times the control value; these two perturbation simulations are hereafter referred to as $0.5\times\mathrm{W}$ and $1.5\times\mathrm{W}$ respectively.
In addition, we run a set of simulations where we apply an additive surface meridional heat flux (anomaly map shown in Fig.~\ref{fig:Model_setup}c) with equal and opposite signed anomalies in the subtropics and subpolar regions (smoothly connected using a hyperbolic tangent function).
This choice of perturbation is not only expected to change the Atlantic Ocean's total MHT, but also the heat carried by each cell.
We analyze four cases (shown in Fig.~\ref{fig:Model_setup}a) where we either reduce or increase the surface meridional heat flux contrast.
The naming of surface heat flux experiments is based on the anomalous surface meridional heat flux contrast applied.
For example, the $-20\,\mathrm{W}\,\mathrm{m}^{-2}$ experiment is constructed by cooling subtropics and warming high latitudes by $10\,\mathrm{W}\,\mathrm{m}^{-2}$ (Fig.~\ref{fig:Model_setup}c).
Each perturbation experiment is run for 100 years to allow advective responses in both the surface and the deep circulations to variations in surface forcing; the first 10 and last 50 years of each experiment are analyzed. 

\subsection{Definition of warm, cold, and mixed cells in latitude--temperature space}
\label{subsec:cell_definitions}

The streamfunction $\Psi$ in latitude--temperature space is defined as (see e.g.,~\cite{Ferrari2011,Holmes2019}):
\begin{equation}
\label{eq:lat-temp-streamfunction}
    \Psi (\phi, \theta, t) \defn R \int_{\lambda_w}^{\lambda_e} \int_{-D}^{\eta} v(\lambda, \phi, z, t) \, \mathcal{H}(\theta - \theta_{\text{c}}(\lambda, \phi, z, t)) \cos{\phi} \, \mathrm{d}z \, \mathrm{d}\lambda,
\end{equation}
where $R$ is the Earth's radius, $\lambda$ is longitude, $\phi$ is latitude, $v$ is meridional velocity, $\lambda_w$ and $\lambda_e$ are the western and eastern extents of the Atlantic basin, $D$ is ocean depth, $\eta$ is sea level, $\mathcal{H}$ is the Heaviside function, and $\theta_{c}$ is conservative temperature \citep{TEOS10}. 
Equation~\eqref{eq:lat-temp-streamfunction} assumes that each temperature level is uniquely related to a depth level.
This assumption is valid in the ocean everywhere except in localized convective regions where temperature inversions can break the unique remapping from depth to temperature.
We compute the streamfunction for $\theta \in [-2, 60] ^{\circ}$C and with a 1$^{\circ}$C interval; recomputing~\eqref{eq:lat-temp-streamfunction} with $0.5^{\circ}$C or $2^{\circ}$C intervals gave similar results.
We use monthly-averaged velocity and temperature diagnostics to compute the streamfunction, neglecting the small contribution of sub-monthly variability to the MHT \citep{Yung2023}.

The streamfunction~\eqref{eq:lat-temp-streamfunction} for the flux-forced control simulation is shown in Fig.~\ref{fig:Model_setup}d. 
Positive values denote clockwise circulation comprising warmer waters flowing northward and colder waters returning southward.
We also observe weak anti-clockwise circulation cells associated with Antarctic Bottom Water at temperatures below $3^\circ$C and the Southern Hemisphere subtropical gyre at warm temperatures.
We divide the clockwise circulation (green colors in Fig.~\ref{fig:Model_setup}d) into warm ($\Psi_{\text{warm}}$), cold ($\Psi_{\text{cold}}$), and mixed ($\Psi_{\text{mixed}}$) cells following \cite{Ferrari2011} (also shown for $26^{\circ}$N in Fig.~\ref{fig:Model_setup}e). 
The warm cell (enclosed by the red contour in Fig.~\ref{fig:Model_setup}d) resembles the subtropical gyre.
Similarly, the cold cell (bounded by the blue contour in Fig.~\ref{fig:Model_setup}d) resembles the AMOC.
A significant proportion of the clockwise circulation transits around and between the warm and cold cells and is captured by the mixed circulation, which is present at all temperatures.
%
The strength of the mixed circulation is quantified as the closed streamfunction contour with the smallest value of $\Psi$ that completely isolates the warm and cold cells. 
This streamfunction contour separates the warm and cold cells at the saddle point (denoted by temperature $\theta_{\text{saddle}}$ and latitude $\phi_{\text{saddle}}$  in Fig.~\ref{fig:Model_setup}d).
After locating the saddle point, we compute the strength of the warm and cold circulations by subtracting the mixed circulation for regions with $\Psi > \Psi_{\text{mixed}}$ (see Fig.~\ref{fig:Model_setup}e) from the streamfunction $\Psi$.
Specifically,
\begin{align}
\label{eq:psiwarm}
    \Psi_{\text{warm}} &\defn \begin{cases}
        \Psi - \Psi_{\text{mixed}} \qquad \text{ if } \Psi > \Psi_{\text{mixed}} \text{ and } \theta \ge \theta_{\text{saddle}}, \\
        0 \qquad \qquad \qquad \qquad \qquad \qquad \text{otherwise,}
    \end{cases}\\
\label{eq:psicold}
    \Psi_{\text{cold}} &\defn \begin{cases}
        \Psi - \Psi_{\text{mixed}} \qquad \text{ if } \Psi > \Psi_{\text{mixed}} \text{ and } \theta < \theta_{\text{saddle}}, \\
        0 \qquad \qquad \qquad \qquad \qquad \qquad \text{otherwise.}
    \end{cases}
\end{align} 
Following these definitions, the flux-forced control streamfunction in Fig.~\ref{fig:Model_setup}d implies that the mixed circulation strength is~7.4$\,$Sv, the warm (gyre) cell strength is~2.1$\,$Sv and the cold (overturning) cell strength is~14.2$\,$Sv.

\subsection{Heatfunction}
\label{subsec:heatfunction}

One advantage of using a streamfunction in temperature--latitude space is that the MHT is easily computed via a cumulative integral in temperature (see e.g.,~\cite{Ferrari2011}). This defines a \textit{heatfunction},
\begin{equation}
\label{eq:heatfunc}
    \mathrm{H}(\phi, \theta, t) \defn \rho_0 c_p \int_{-2^{\circ}\text{C}}^{\theta} \Psi (\phi, \theta', t) \, \mathrm{d}\theta' ,
\end{equation}
where $\rho_0$ is reference density and $c_p$ the specific heat capacity (both constants in our calculations since we use conservative temperature).
The heatfunction~\eqref{eq:heatfunc} is equal to the cumulative MHT for waters colder than $\theta$, and estimates the MHT of a closed circulation (in which mass is conserved) at a given latitude $\phi$ as:
\begin{equation}
\label{eq:MHT}
    \mathrm{MHT}(\phi) = \mathrm{H}(\phi, \theta^{\text{max}}) - \mathrm{H}(\phi, \theta^{\text{min}}),
\end{equation}
where $\theta^{\text{min}}$ and $\theta^{\text{max}}$ are the minimum and maximum temperatures spanned by the circulation cell. From \eqref{eq:heatfunc} and \eqref{eq:MHT}, the MHT at a given latitude for each cell is directly proportional to its circulation strength and the temperature range spanned by that circulation \citep{Ferrari2011}.

The heatfunction for the flux-forced control simulation (Fig.~\ref{fig:Model_setup}f) is mostly positive and increases with temperature, demonstrating the dominant northward heat transport throughout the Atlantic associated with the clockwise circulation in the latitude--temperature plane.
The total northward MHT, indicated by the value of the heatfunction at the maximum temperature, is largest in the subtropics, where clockwise circulation spans the largest temperature range.
Heatfunction contours that pierce the ocean's surface at latitudes north of $35^\circ$N indicate heat loss to the atmosphere \citep{Holmes2019}. 
At steady state, the horizontal divergence of heatfunction contours quantify the air-sea heat exchange.

\section{Results}
\label{sec:Results}
 
The MHT due to each cell, computed using~\eqref{eq:MHT}, can vary between the simulations via:~\emph{(i)}~adjustments in the circulation strength $\Psi$, \emph{(ii)}~modifications in the temperature range spanned by the circulation, or \emph{(iii)}~a combination of both of the above. 
These processes occur on different timescales, thus, we examine the heat carried meridionally by each cell on short (1--10 years) and long (51--100 years) timescales.

\subsection{Wind stress perturbation experiments}
\label{subsec:winds}

We analyze two perturbation experiments in which we change wind stress in the North Atlantic by a factor of 0.5 and 1.5 (labeled as 0.5$\times$W and 1.5$\times$W respectively) times the control (CTRL) value. 

\subsubsection{Initial response}

\begin{figure}
    \includegraphics[width=1.0\textwidth]{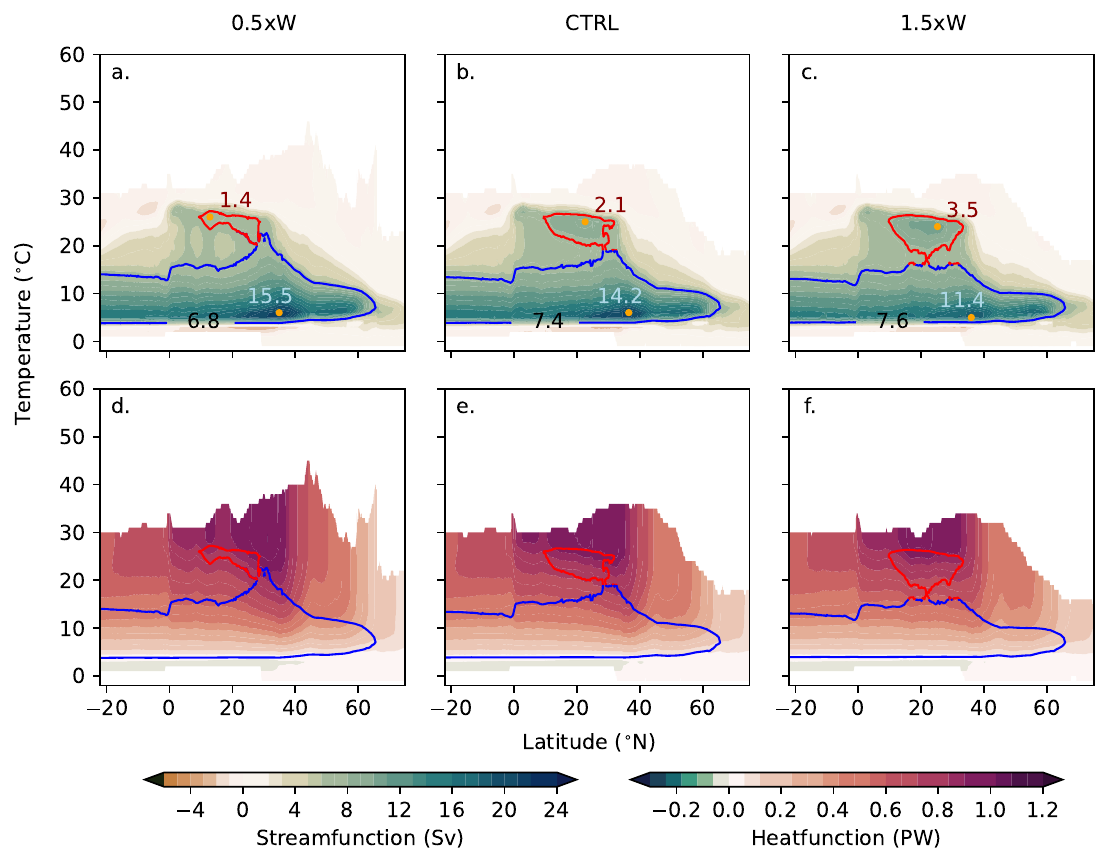}\vspace{-1em}
    \caption{(a-c). Streamfunction in latitude--temperature space (left colorbar) calculated from the first 10 years for the 0.5$\times$W (left), CTRL (middle), and 1.5$\times$W (right) experiments. 
    The strength of the mixed cell (black text), the warm cell (red text), and the cold cell (blue text) are indicated. 
    (d-f). Heatfunction (right colorbar) averaged over the first 10 years for the 0.5$\times$W (left), CTRL (middle), and 1.5$\times$W (right) experiments.}
    \label{fig:Wind_1-10}
\end{figure}

The circulation strength at warm temperatures -- that is supported by contributions from both warm and mixed cells -- scales with the magnitude of wind stress on short timescales (compare streamfunction within the red contour between Figs.~\ref{fig:Wind_1-10}a-c). 
These changes are primarily led by the warm cell, with a smaller contribution from the mixed cell (compare trends in the red and black text across experiments in Figs.~\ref{fig:Wind_1-10}a-c).
For example, the circulation strength of the warm cell changes by 150\% between $0.5\times$W and $1.5\times$W compared to a relatively modest $11.7\%$ change in the mixed cell.
The core of the circulation anomalies at warm temperatures resides between $20^{\circ}\mathrm{N}-40^{\circ}\mathrm{N}$, likely because increased wind stress upwells a larger fraction of the northward flow and feeds it back to the warm and mixed cells via southward Ekman transport.
The increased fraction of southward Ekman flow at warm temperatures also explains the reduction in the circulation strength at cold temperatures at those latitudes (see trends in the cold text between Figs.~\ref{fig:Wind_1-10}a-c), since less waters likely propagate poleward in the $1.5\times$W simulation to circulate within the cold cell.

The heat carried meridionally by each cell is determined by its circulation strength and the temperature difference between its northward and southward flowing branches (see Eqs.~\eqref{eq:heatfunc} and~\eqref{eq:MHT}).
For the warm cell, both circulation and temperature contrast scale with wind stress during the first 10 years. 
As a result, its MHT increases with wind stress -- for example, from $1.44\times10^{13}\,\mathrm{W}$ in the $0.5\times$W simulation to $5.91\times10^{13}\,\mathrm{W}$ in the $1.5\times$W simulation at the latitudes of peak circulation (marked by red dots in Figs.~\ref{fig:Wind_1-10}a-c).
In contrast, circulation-driven MHT anomalies in the mixed cell are partially compensated by temperature distribution–driven anomalies, resulting in an increase in the MHT with the wind stress.
For the cold cell, its MHT at the latitude of peak circulation decreases from $4.07\times10^{14}\,\mathrm{W}$ to $2.32\times10^{14}\,\mathrm{W}$ between the $0.5\times$W and $1.5\times$W simulations.
A reduction in the cold cell's MHT in the $1.5\times\mathrm{W}$ experiment leads to cooler waters north of 35$^{\circ}$N, while weaker MHT in the $0.5\times$W case results in warmer conditions. 
Similarly, stronger northward heat transport by the warm cell in $1.5\times$W contributes to cooling the tropics, while weaker MHT in the $0.5\times$W leads to warmer tropics.


\subsubsection{Long-term response}

\begin{figure}
    \includegraphics[width=1.0\textwidth]{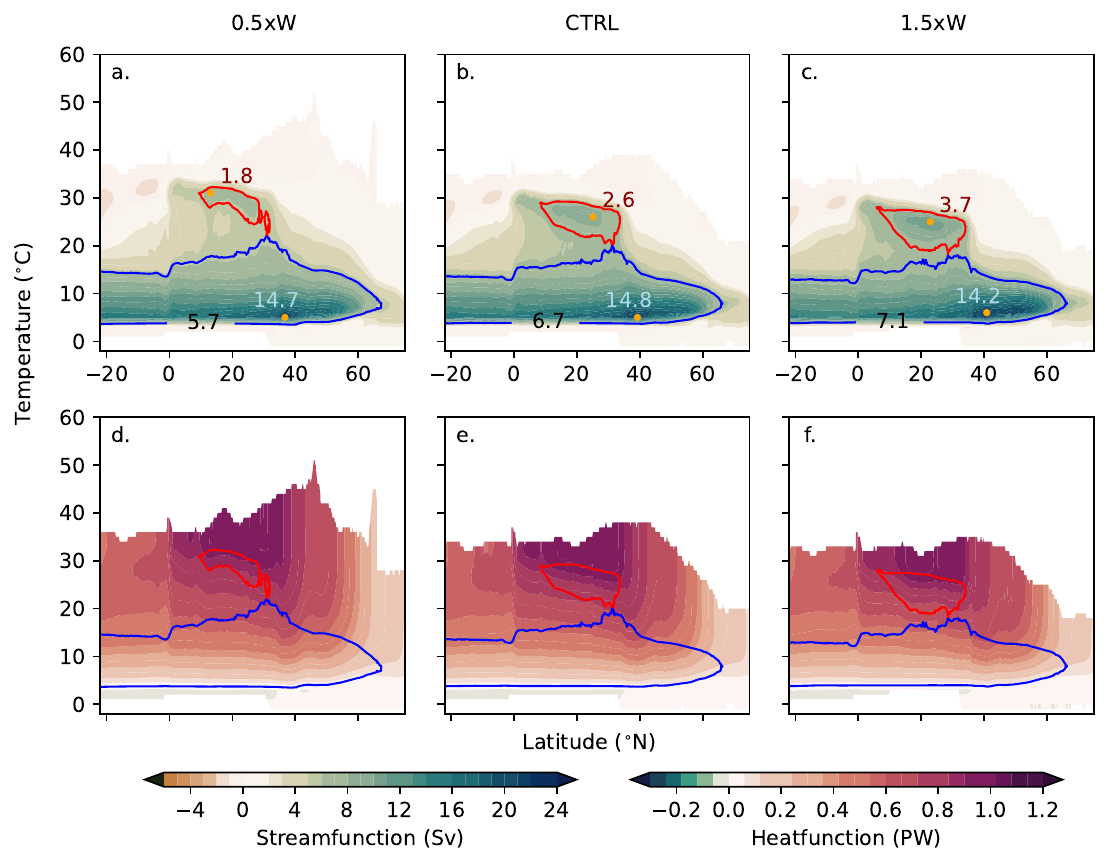}\vspace{-1em}
    \caption{(a-c). Streamfunction in latitude--temperature space (left colorbar) calculated from the last 50 years for the 0.5$\times$W (left), CTRL (middle), and 1.5$\times$W (right) experiments. 
    The strength of the mixed cell (black text), the warm cell (red text), and the cold cell (blue text) are indicated. 
    (d-f). Heatfunction (right colorbar) averaged over the last 50 years for the 0.5$\times$W (left), CTRL (middle), and 1.5$\times$W (right) experiments.}
    \label{fig:Wind_51-100}
\end{figure}

The warm and mixed cells' circulation anomalies in the last 50 years do not change significantly between experiments compared to the first 10 years, suggesting that the two cells quickly adjust to the changes in wind stress.
The cold cell's circulation is minutely affected by Atlantic-only wind stress perturbations on long timescales (compare trends in the blue text between Figs.~\ref{fig:Wind_51-100}a-c).
Conversely, experiments with global wind stress perturbations (not shown) produce a direct scaling between the cold cell's circulation strength and the magnitude of wind stress, which suggests that anomalous wind-driven upwelling in the Southern Ocean plays a key role in modulating the cold cell \citep{Cessi2019}.
Variations in the circulation-driven MHT thermally restructure each cell.
For example, the temperature distribution anomalies observed in the tropics and high-latitudes in the first 10 years (Figs.~\ref{fig:Wind_1-10}a-c) are amplified in the last 50 years (Figs.~\ref{fig:Wind_51-100}a-c).

On long timescales, modifications in the ocean's temperature partially compensate for circulation-driven MHT changes (compare Figs.~\ref{fig:Wind_51-100}d-f) for the mixed cell.
Nevertheless, the circulation anomalies dominate, and the mixed cell's MHT increases from $6.57\times10^{14}\,\textrm{W}$ in the $0.5\times$W simulation to $7.00\times10^{14}\,\textrm{W}$ in the $1.5\times$W simulation.
In contrast, anomalies in both the circulation strength and temperature distribution act jointly in varying the MHT by the warm and cold cells.
The warm cell carries more heat as the wind stress strengthens while the cold cell's MHT reduces as the wind stress intensifies (compare Figs.~\ref{fig:Wind_51-100}d-f).
However, the total Atlantic MHT remains the same since the total surface heat fluxes are fixed across the three experiments. 
Such compensation in the MHT anomalies between the three cells is not observed in the wind perturbation experiments by \cite{Ferrari2011}, since their surface buoyancy forcing varies with different values of wind stress.
To summarise, the total MHT is unchanged in our experiments on long timescales, but the proportion of heat carried by the warm and mixed cells increases with wind stress at the expense of the cold cell.

\subsection{Surface buoyancy flux contrast experiments}
\label{subsec:heat}

We now examine the response of each cell's circulation and heat transport to an increase and reduction in the North Atlantic surface meridional heat flux contrast, as quantified by the $-20\,\textrm{W}\,\textrm{m}^2$ and $+20\,\textrm{W}\,\textrm{m}^2$ simulations.
%
%
Since we apply anomalous surface meridional heat flux gradients, we expect variations in the total Atlantic MHT.
For example, in the reduced surface meridional buoyancy flux contrast experiments we expect a reduction in the total MHT.

\subsubsection{Initial response}
\begin{figure}
    \includegraphics[width=1.0\textwidth]{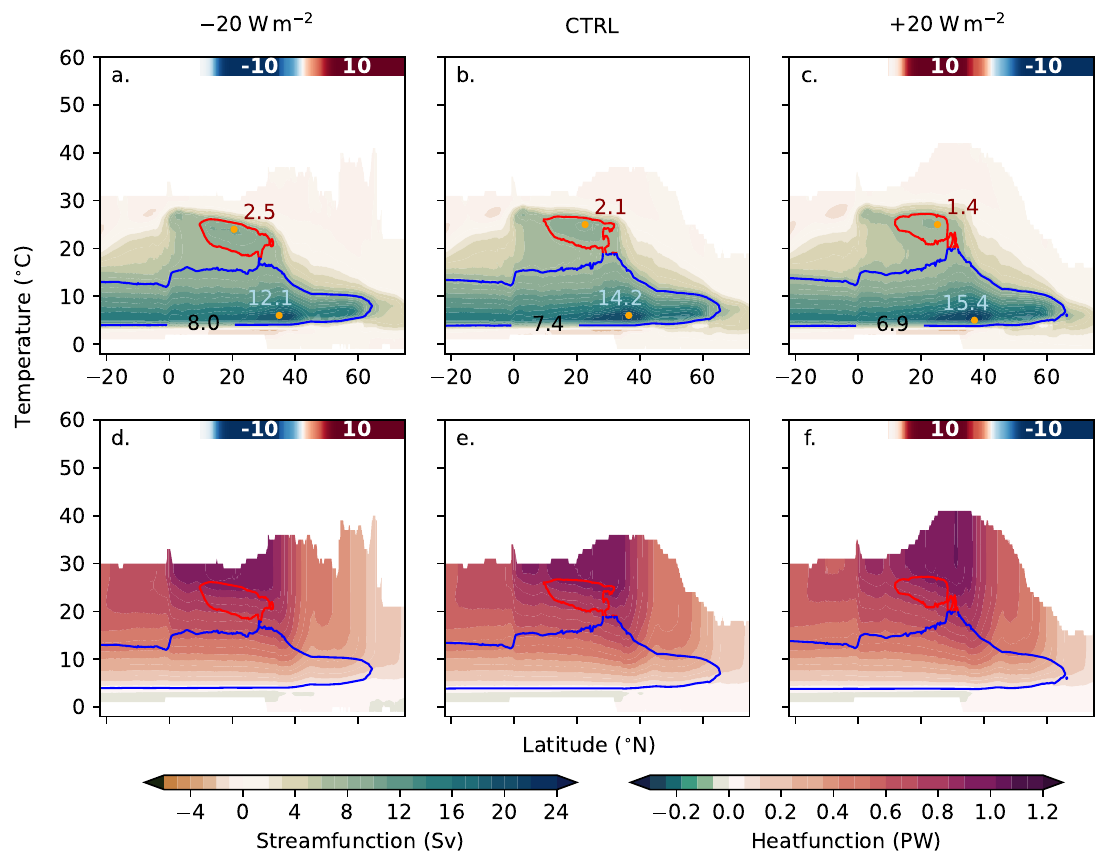}\vspace{-1em}
    \caption{(a-c). Streamfunction in latitude--temperature space from the first 10 years for the $-20 \,\mathrm{W}\,\mathrm{m}^{-2}$ (left column), CTRL (middle column), and $+20 \,\mathrm{W}\,\mathrm{m}^{-2}$ (right column) experiments. 
    The strength of the mixed cell (black text), the warm cell (red text), and the cold cell (blue text) are indicated.
    The colorbar at the top of (a), (c), (d), and~(f) indicates the anomalous surface heat fluxes (in $\mathrm{W}\,\mathrm{m}^{-2}$).
    (d-f). Heatfunction (right colorbar) averaged over the first 10 years for the surface meridional heat flux contrast perturbation experiments.}
    \label{fig:Heat_1-10}
\end{figure}

In the first 10 years, the cold cell's circulation strengthens by 27.3\%  between the $-20\,\mathrm{W}\,\mathrm{m}^{-2}$ and $+20\,\mathrm{W}\,\mathrm{m}^{-2}$ simulations (compare blue text between Figs.~\ref{fig:Heat_1-10}a-c).
The core of the circulation anomalies is located northward of $35^{\circ}$N, which occurs due to anomalous surface heat loss in the western boundary region and in the subpolar regions.
This anomalous surface heat loss enhances deep water formation and strengthens the cold cell in the $+20\,\mathrm{W}\,\mathrm{m}^{-2}$ simulation, whereas reduced surface heat loss weakens the cold cell's circulation in the $-20\,\mathrm{W}\,\mathrm{m}^{-2}$ simulation.
The warm and mixed cells weaken in the first 10 years as the surface meridional heat flux contrast increases (compare black text between Figs.~\ref{fig:Heat_1-10}a-c).

The cold cell's MHT at the latitudes of the peak circulation increases from $3.11\times10^{14}\,\mathrm{W}$ to $3.58\times10^{14}\,\mathrm{W}$ between the $-20\,\mathrm{W}\,\mathrm{m}^{-2}$ and $+20\,\mathrm{W}\,\mathrm{m}^{-2}$ simulations (compare heatfunction plots between Figs.~\ref{fig:Heat_1-10}d-f).
Unlike the wind stress perturbation experiments, modifications in the cold cell's MHT stem primarily from changes in the circulation in the first 10 years. 
In addition, the warm and mixed cells' MHT reduce as the surface meridional heat flux contrast increases, which also occur primarily due to modifications in the strength of the two cells' circulations.
We note that changes in the temperature distribution in the first 10 years are largely affected by localized surface heat flux anomalies rather than by circulation-driven MHT variations.
For example, anomalous surface heating raises the maximum temperature in subpolar regions in the $-20\,\mathrm{W}\,\mathrm{m}^{-2}$ experiment and in the subtropics in the $+20\,\mathrm{W}\,\mathrm{m}^{-2}$ experiment (Figs.~\ref{fig:Heat_1-10}d-f).

\subsubsection{Long-term response}

\begin{figure}
    \includegraphics[width=1.0\textwidth]{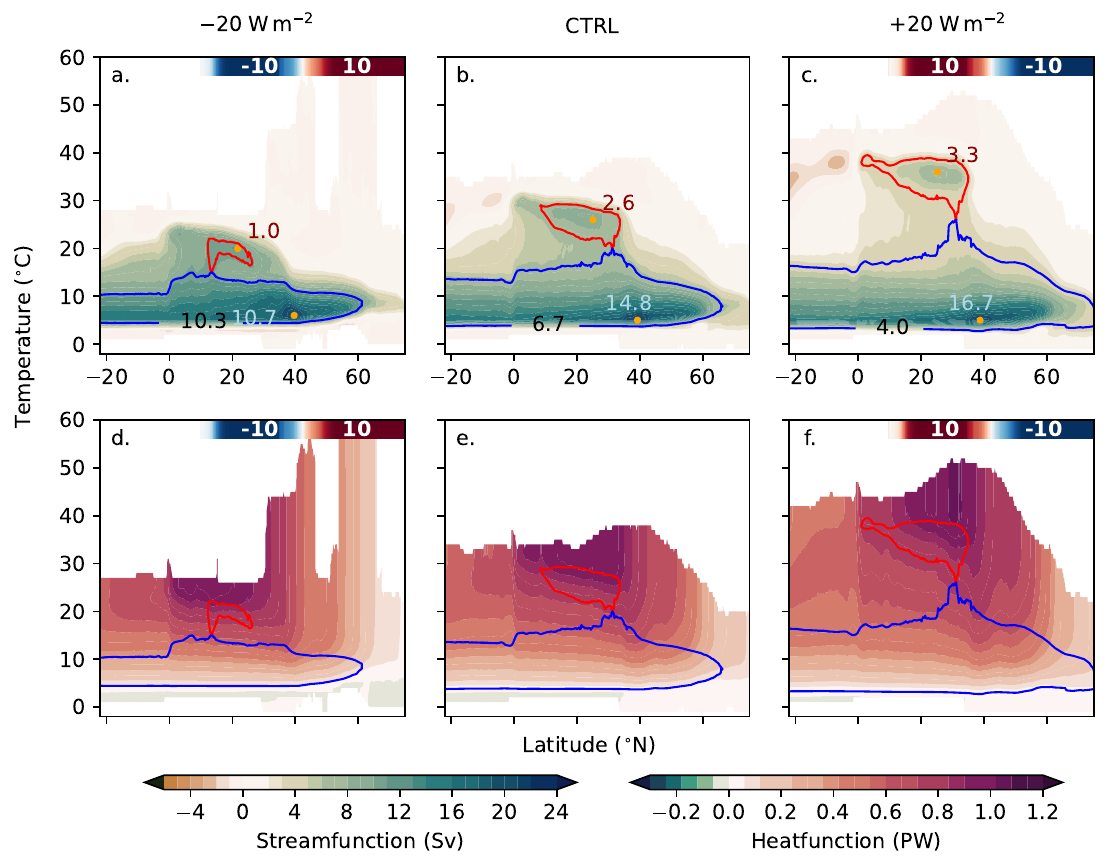}\vspace{-1em}
    \caption{(a-c). Streamfunction in latitude--temperature space from the last 50 years for the $-20\,\mathrm{W}\,\mathrm{m}^{-2}$ (left column), CTRL (middle column), and $+20 \,\mathrm{W}\,\mathrm{m}^{-2}$ (right column) experiments. 
    The strength of the mixed cell (black text), the warm cell (red text), and the cold cell (blue text) are indicated.
    The colorbar at the top of (a), (c), (d), and~(f) indicates the anomalous surface heat fluxes (in $\mathrm{W}\,\mathrm{m}^{-2}$).
    (d-f). Heatfunction (right colorbar) averaged over the last 50 years for the surface meridional heat flux contrast perturbation experiments.}
    \label{fig:Heat_51-100}
\end{figure}

The trends in circulation and temperature distribution observed within each cell in the first 10 years strengthen in the last 50 years.
The warm and cold cells' circulations increase by 230\% and 56.1\% respectively between $-20\,\mathrm{W}\,\mathrm{m}^{-2}$ and $+20\,\mathrm{W}\,\mathrm{m}^{-2}$ simulations.
In contrast, the mixed cell's circulation reduces by 61.1\% between $-20\,\mathrm{W}\,\mathrm{m}^{-2}$ and $+20\,\mathrm{W}\,\mathrm{m}^{-2}$ simulations (compare streamfunction plots between Figs.~\ref{fig:Heat_51-100}a-c).
It is interesting that the warm cell's circulation strength shows a trend reversal compared to the first 10 years.
The long term trend is consistent with recent studies (e.g., see \cite{Hogg2020} and \cite{Bhagtani2023}) demonstrating that the near-surface circulation scales with the surface meridional heat flux contrast.
However, we note that we define the gyre as waters warmer than a given temperature, while past studies use different definitions (e.g., the maximum of the barotropic streamfunction in the subtropics by \cite{Hogg2020} or waters lighter than a given density class by \cite{Bhagtani2023}).

Similar to the wind stress perturbation experiments, the circulation-driven anomalies feed back onto the temperature structure on long timescales.
For each cell, we observe a direct relationship between the temperature range spanned by the circulation and the surface meridional heat flux contrast (compare heatfunction plots between Figs.~\ref{fig:Heat_51-100}d-f).
Furthermore, modifications in each cell's temperature distribution occurs across all latitudes (Figs.~\ref{fig:Heat_51-100}d-f). 
For example, the maximum temperature reduces in the $-20\,\mathrm{W}\,\mathrm{m}^{-2}$ experiment and increases in the $+20\,\mathrm{W}\,\mathrm{m}^{-2}$ experiment even for latitudes south of $10^{\circ}$N where no anomalous surface heat flux is applied.
This analysis underscores the importance of modifications in the circulation in setting remote changes in the temperature distribution within each cell on long timescales.

We find that variations in the warm and cold cells' strength and temperature distribution act jointly to influence the Atlantic MHT on long timescales (compare the heatfunction plots between Figs.~\ref{fig:Heat_51-100}d-f).
Modifications in the Atlantic Ocean's MHT to the surface meridional heat flux contrast is primarily controlled by the cold cell, with relatively smaller influence by the warm cell.
For example, between the $-20 \,\mathrm{W}\,\mathrm{m}^{-2}$ and $+20 \,\mathrm{W}\,\mathrm{m}^{-2}$ experiments, the heat carried northward by the cold cell increases from $2.18\times10^{14}\,\mathrm{W}$ to $4.44\times10^{14}\,\mathrm{W}$, while the heat carried northward by the warm cell increases from $7.37\times10^{12}\,\mathrm{W}$ to $6.49\times10^{13}\,\mathrm{W}$.
In contrast to the warm and cold cells, variations in the temperature distribution and circulation strength compensate the mixed cell's MHT anomalies to a large extent.
More specifically, the mixed cell weakens but occupies a larger temperature range as the surface meridional heat flux contrast increases.
Variations in the circulation dominate over the temperature distribution anomalies; thus, the mixed cell's MHT reduces as the surface heat flux contrast increases.
Overall, as expected, the total Atlantic MHT scales with the surface meridional heat flux contrast (Figs.~\ref{fig:Heat_51-100}d-f).


\section{Summary and discussion}
\label{sec:discussion}

In this study, we use a series of global ocean simulations to quantify the independent effects of wind stress and surface heat fluxes in steering the North Atlantic MHT. 
We divide the Atlantic Ocean's circulation into a warm and cold cell (resembling a gyre and mid-depth overturning circulation respectively), and a mixed circulation to highlight water masses transitioning between the warm and mixed cells. 
This division is performed in latitude--temperature space following \cite{Ferrari2011} to understand the sensitivity of each circulation's MHT to varying surface forcing (Figs.~\ref{fig:Model_setup}d-e). 
The MHT of the warm, cold, and mixed cells adjust to changes in winds and surface heat fluxes through three mechanisms: \emph{(i)}~changes in circulation strength, \emph{(ii)} variations in the temperature distribution of each cell (i.e., the average temperature difference between the northward and southward flowing arms of the cell), or \emph{(iii)}~a combination of both.
We find that the anomalies in circulation and temperature distribution due to changes in surface forcing act jointly for the warm and cold cells, whereas they compensate each other for the mixed cell.

\begin{figure}[!ht]
    \includegraphics[width=1.0\textwidth]{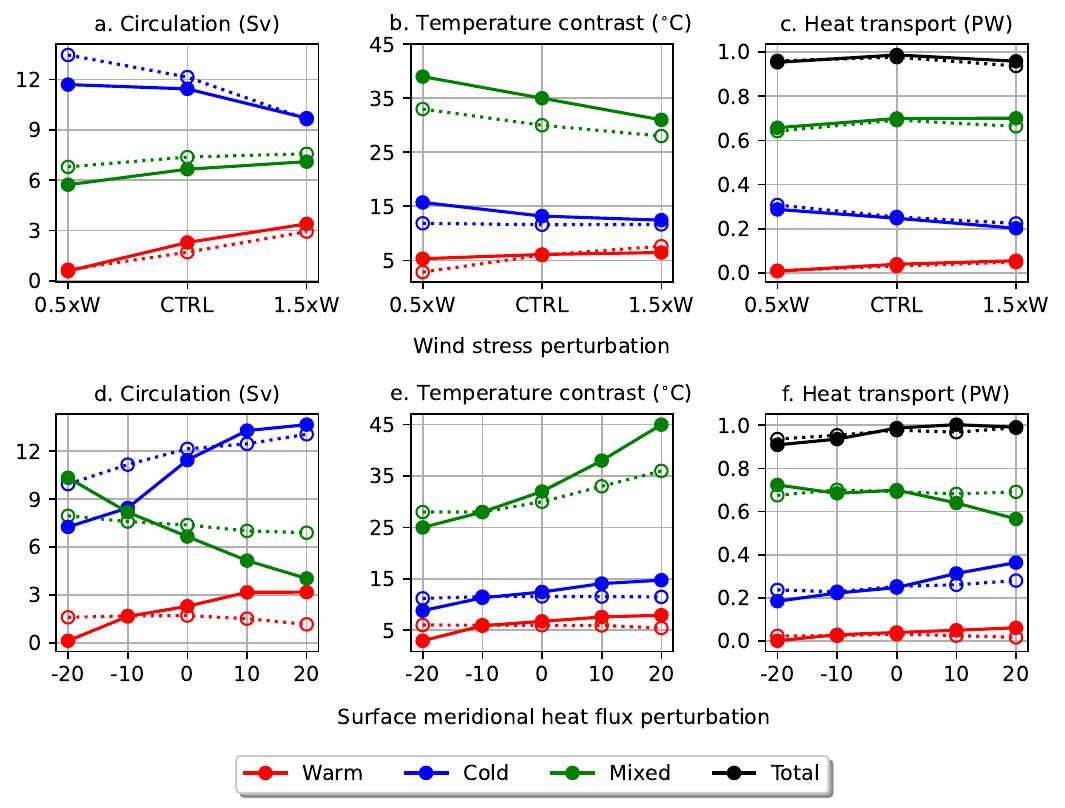}
    \caption{Metrics for meridional heat transport for the first 10 years (dashed lines) and last 50 years (solid lines) for the wind perturbation (top row) and surface meridional heat flux perturbation (bottom row) experiments.}
    \label{fig:Wind_Heat_contrast_MHT_metrics}
\end{figure}

%
We select the $26^\circ$N transect to summarize changes in the circulation and the MHT for each cell in each set of perturbation experiments. 
Contrary to the classical literature, we find more interconnected roles of wind stress and surface buoyancy forcing in steering the MHT by the near-surface and deep circulations (Figs.~\ref{fig:Wind_Heat_contrast_MHT_metrics}c,~f). 
In the last 50 years, the warm and mixed cell's MHT increase with the strength of wind stress.
The total MHT in these wind stress perturbation experiments (in which we maintain the same surface buoyancy fluxes) should converge to the CTRL's MHT as we progress towards equilibrium.
We find that this is achieved through a reduction in the cold cell's MHT that compensates the increased heat carried by the warm and mixed cells on long timescales. 

Anomalies in the MHT due to changes in the surface meridional heat flux contrast are primarily attributed to the cold cell, with partial compensation by the mixed cell (blue and green lines in Figs.~\ref{fig:Wind_Heat_contrast_MHT_metrics}f).
The warm cell's MHT also scales with the surface meridional heat flux contrast (red line in Fig.~\ref{fig:Wind_Heat_contrast_MHT_metrics}d).
The scaling seen in the warm and cold cells is the result of a direct relationship between the strength of the warm cell's circulation and temperature range with the surface meridional heat flux contrast.
The trend in the warm cell's circulation is consistent with recent studies \citep{Hogg2020, Bhagtani2023}, thereby demonstrating an important role of surface buoyancy forcing in steering the gyre circulation.
The total Atlantic Ocean's MHT scales with the surface meridional heat flux contrast.
In conclusion, our analysis underscores the distinct pathways utilized by the three cells to respond to varying surface forcing.

In reality, the wind stress and surface buoyancy forcing are strongly coupled and depend on the ocean response itself.
In particular, the surface heat flux depends strongly on the ocean's sea surface temperature, which responds dynamically to heat transport changes associated with both winds and surface heat fluxes. 
In the absence of a restoring, the flux-forced simulations show a higher SST variance that is partly attributed to reduced damping of mesoscale eddies, but also partly caused by interannual-to-multidecadal variability in the North Atlantic. 
However, since the ocean's MHT is majorly driven by the time-mean MHT \citep{Yung2023}, the higher SST variance does not affect our results to leading order.

Our flux-forced simulations, along with a decomposition of the Atlantic circulation into warm, cold, and mixed cells, provide useful insights into how the Atlantic Ocean's MHT may respond to climate change.
For example, climate projections suggest a strengthening of zonal winds \citep{Shaw2024}. 
Under such scenarios, our results indicate an increased contribution of the warm cell to northward heat transport, led by a spin up of the subtropical gyre.
At the same time, projected melting in the Greenland and the Arctic \citep{Rantanen2022} will likely reduce surface buoyancy loss in the North Atlantic leading to a reduction in the MHT carried by the AMOC.
Under either scenario, the mixed circulation is expected to intensify, implying \emph{(i)}~a stronger coupling between near-surface and deep circulations, and \emph{(ii)}~a larger proportion of the Atlantic Ocean's MHT being carried by water masses that can transit between the subtropical gyre and the AMOC.

\section*{Acknowledgments}

We thank Christopher Bladwell, William Johns, Spencer Jones, and the Consortium for Ocean--Sea Ice Modeling in Australia (\href{https://cosima.org.au}{cosima.org.au}) for fruitful discussions and for maintaining the Cookbook of analysis recipes \citep{cosima-cookbook}.
Computational resources were provided by the National Computational Infrastructure, which is supported by the Commonwealth Government of Australia.

\section*{Data Availability}

Notebooks used for reproducing the analyses and figures are available at the repository \href{https://github.com/dhruvbhagtani/Atlantic-meridional-heat-transport-heatfunction}
{github.com/dhruvbhagtani/Atlantic-meridional-heat-transport-heatfunction}.
Preprocessed data for the flux-forced simulations is available online at \cite{Bhagtani_2023_8405009}.
The modified MOM5 source code for the flux-forced simulations used in this paper is available at \href{https://github.com/dhruvbhagtani/MOM5}{github.com/dhruvbhagtani/MOM5}.
Our analyses were facilitated with the Python packages \texttt{dask} \citep{Rocklin2015} and \texttt{xarray} \citep{Hoyer2017}.

\section*{Funding}
We acknowledge funding from the Australian National University under the University Research Scholarship~(D.B.) and from the Australian Research Council under DECRA Fellowships DE210100004~(R.M.H.) and DE210100749~(N.C.C.).

\section*{Conflict of Interest}
The authors declare no conflict of interest relevant to this study.





\bibliography{sn-bibliography}

\end{document}